# Empirical Evaluation of the Segment Anything Model (SAM) for Brain Tumor Segmentation

Authors: Mohammad Peivandi[*,1], Jason Zhang[*,2], Michael Lu[*,3], Dongxiao Zhu[4], Zhifeng Kou[5]

1 - Department of Biomedical Engineering, Wayne State University, Detroit, MI, United States

2 - Canton High School, Canton, MI, United States

3 - Phillips Exeter Academy, Exeter, NH, United States

4 - Department of Computer Science, Wayne State University, Detroit, MI, United States

5 - Department of Biomedical Engineering, Wayne State University, Detroit, MI, United States

*Equal contribution.

## Abstract

Brain tumor segmentation presents a formidable challenge in the field of Medical Image Segmentation. While deep-learning models have been useful, human expert segmentation remains the most accurate method. The recently released Segment Anything Model (SAM) has opened up the opportunity to apply foundation models to this difficult task. However, SAM was primarily trained on diverse natural images. This makes applying SAM to biomedical segmentation, such as brain tumors with less defined boundaries, challenging.

In this paper, we enhanced SAM's mask decoder using transfer learning with the Decathlon brain tumor dataset. We developed three methods to encapsulate the four-dimensional data into three dimensions for SAM. An on-the-fly data augmentation approach has been used with a combination of rotations and elastic deformations to increase the size of the training dataset. Two key metrics: the Dice Similarity Coefficient (DSC) and the Hausdorff Distance 95th Percentile (HD95), have been applied to assess the performance of our segmentation models. These metrics provided valuable insights into the quality of the segmentation results. In our evaluation, we compared this improved model to two benchmarks: the pretrained SAM and the widely used model, nnUNetv2. We find that the improved SAM shows considerable improvement over the pretrained SAM, while nnUNetv2 outperformed the improved SAM in terms of overall segmentation accuracy. Nevertheless, the improved SAM demonstrated slightly more consistent results than nnUNetv2, especially on challenging cases that can lead to larger Hausdorff distances. In the future, more advanced techniques can be applied in order to further improve the performance of SAM on brain tumor segmentation.

# 1. Introduction

The application of deep learning models to Medical Image Segmentation is a fast-growing research domain. In particular, brain tumor segmentation is a difficult task, even for trained experts. Neuroscientist experts have achieved a cross-rater Dice score of 0.77-0.93 for whole tumor segmentation (core+edema), showing the challenges of brain tumor segmentation [1]. Although manual segmentation is still the most accurate method of identifying certain regions, it is laborious and time-intensive. Therefore, there is a high demand for models that can quickly and accurately segment brain tumors in the medical field.

The recent release of the Segment Anything Model (SAM) has introduced new opportunities for researchers to experiment with zero-shot semantic segmentation techniques in the field of medical image segmentation. Although SAM has strong generalization to general fields, having been trained on a wide variety of natural images, M. Ahmadi et. [2] and C. Mattjie et. al [3] shows that SAM requires adjustments in order to perform well on medical image segmentation.

The ability to segment different brain tumor Region of Interests (ROIs), or tumor subregions, holds significant clinical value. Different ROIs can help identify the type and stage of a tumor. For example, the existence of an enhancing region can differentiate between certain types of gliomas. Segmenting different ROIs also helps with the classification of tumor type, which then influences treatment strategies by medical doctors. Consequently, fine-tuned SAM needs to be evaluated on its ability to segment different subregions of brain tumors in order to accurately assess its usefulness.

Segmenting the whole tumor (combined regions of enhancing, non-enhancing, and edema) is also crucial for many clinical treatments. For instance, assessing the whole tumor is crucial for surgical planning, determining the feasibility of surgical removal, and estimating the potential risks and benefits of the surgery.

Jun Ma et al. [4], in their work on MedSAM, utilized an extensive dataset of over one million medical images spanning 15 imaging modalities, showcasing that MedSAM can be competitive or even surpass widely used models in certain tasks. However, they encountered challenges with modality imbalance and experienced poor performance in tasks such as the segmentation of vessel-like branching structures. Specifically, for brain tumor segmentation, they relied solely on one modality and did not delve into a detailed analysis of the model. In alignment with their approach, we fine-tune the mask decoder of SAM, drawing inspiration from their technique. Our findings reveal that this fine-tuning not only enhances the SAM architecture's capabilities but also positions it as a strong competitor against other widely-used models.

## 2. Related works:

The evolution of deep learning techniques for medical image segmentation has seen remarkable advancements over the recent years. The U-Net architecture, introduced by Ronneberger et al. [5], emerged as a pioneering model specifically designed for medical image segmentation. Its symmetric encoder-decoder structure and skip connections facilitate precise localization, making it a benchmark for many medical imaging tasks. However, with the advent of transformer architectures, which initially revolutionized natural language processing, there has been an interest towards their application in computer vision tasks. Dosovitskiy et al. (2020) [6] introduced the Vision Transformer (ViT) that divides images into fixed-size patches, linearly embeds them, and then processes them with transformer encoders. Based on this structure, the Segment Anything Model (SAM) by Meta was proposed as a zero-shot semantic segmentation model, leveraging the capabilities of transformers. Recent studies have started to explore the potential of SAM for medical imaging, showing a promising direction for future research in achieving better segmentation performance.

Kaidong Zhang, Dong Liu [7] proposed SAMed that applies the low-rank-based (LoRA) fine-tuning strategy to the SAM image encoder. SAMed is finetuned together with the prompt encoder and the mask decoder and they achieved 81.88 DSC and 20.64 HD on Synapse multi-organ segmentation dataset. SAMed is a very computationally expensive approach as it requires finetuning the SAM image encoder.

Ravi Bhushan Bhardwaj and Dr. Anum Haneef [8] experimented with using SAM and MedSAM on colored retinal fungus images. Using the colored images, they were able to achieve DSC results of 85.97 and 90.15 for a fine-tuned SAM and MedSAM respectively. This is still experimental, but their results show promise for the use of SAM in medical segmentation.

Mohsen Ahmadi et. al [2] compared pre-trained SAM to other widely used models on breast tumor ultrasound and mammography images. They found that pre-trained SAM was heavily reliant on human input in order to perform well. This further demonstrates the need to fine-tune SAM for better performance.

Chen et al. (2023) [9] proposed a ladder fine-tuning approach for SAM, integrating a complementary network for medical image segmentation. The approach consists of two stages: (1) fine-tuning the SAM image encoder with a low-rank approximation (LoRA) strategy similar to SAMed, and (2) fine-tuning the SAM decoder together with a complementary CNN network. The authors evaluated the approach on the Synapse multi-organ segmentation dataset and achieved good results. However, similar to SAMed, this model is computationally expensive as it needs to fine-tune two networks at the same time.

# 3. Materials and Methods:

## 3.1. Segment Anything Model:

The Segment Anything Model (SAM), based on the Vision Transformer (ViT) structure, was trained on the SA-1B dataset with diverse natural images, making it less suitable for biomedical segmentation due to contrasting image features. Its performance on brain tumor segmentation improves significantly after fine-tuning. SAM's architecture, with deep layers, employs text prompts, multi-head self-attention mechanisms, and feed-forward networks in its transformer blocks. The mask decoder refines segmentation using both image and prompt embeddings.

## 3.2. Stochastic Weight Averaging (SWA):

SWA offers a deviation from traditional SGD by computing an average of model weights alongside an adaptive learning rate. This encourages broader optimal solutions and enhanced generalization. Unlike conventional SGD that converges to peripheral solutions, SWA aims for the center of a flat loss region. [10]

## 3.3. Learning Rate Finder

Using PyTorch Lightning's tool, we determined the optimal learning rate for our model, adjusting it every 5 epochs based on the tool's suggestions. [11]

## 3.4. Data Augmentation:

Deep learning models require extensive training data, yet there's a scarcity of publicly available brain tumor images due to patient privacy and segmentation challenges. To address this, we employed data augmentation, specifically using rotations and elastic deformations. While methods like adding random noise, Mixup, and using GAN models exist, our choice was influenced by Goceri [12], who demonstrated the efficacy of simpler augmentations like rotations. It's crucial to acknowledge the brain's asymmetry; thus, flipping images might not yield realistic data, and excessive modifications could distort authenticity. We applied rotations between -20º and 20º to simulate varied head positions in MRI scans and used elastic deformation to account for diverse tumor shapes, enhancing the model's recognition capability in novel images.

## 3.5. 2D-PCA:

2D-Principal Component Analysis (2D-PCA) modifies traditional PCA for matrix or image data. Instead of vectorizing images like standard PCA, 2D-PCA projects image matrices onto orthogonal axes by analyzing covariance for rows and columns. Eigenvalues and eigenvectors derived from these inform the reduced image representation, maintaining primary features and spatial context. While 2D-PCA may not capture all details in multi-modality MRI data or account for inter-modality correlations, we assessed its performance in our study.

## 3.6. Evaluation Metrics:

We used the Dice Similarity Coefficient (DSC) and the Hausdorff Distance (HD95) to gauge our segmentation models' performance.

DSC [13] measures the overlap between predicted segmentation and ground truth in medical imaging. A score of 1 signifies perfect overlap, while 0 means no overlap. Mathematically:

$$DSC(X, Y) = (2 * |X \cap Y|) / (|X| + |Y|) \tag{1}$$

where X and Y represent ground truth and segmentation pixels.

Hausdorff Distance in equation (2) quantifies the maximum distance between boundaries of ground truth and segmentation. While DSC assesses area overlap, HD evaluates boundary alignment. The HD95 version focuses on the smallest 95% of distances, mitigating outlier impact, specifically when most of the segmentation closely matches the ground truth.

$$HD(X,Y) = max\{\sup_{x \in X} \inf_{y \in Y} d(x,y), \sup_{y \in Y} \inf_{x \in X} d(x,y)\} \tag{2}$$

### 3.7. Dataset:

In our research we used the Decathlon dataset, having the FLAIR, T1w, T1-Gd, and T2w modalities, along with corresponding ground truth labels delineating three ROIs: edema, non-enhancing, and enhancing regions. The dataset, containing 484 3D multimodal brain images, was split into 80% training and 20% testing.

We created labels that focus on only one of the ROIs for binary segmentation. They each contain the image of the brain and the ground truth of the brain tumor. On-the-fly data augmentation was used to help with overfitting.

For testing, we take 2D slices along the axial plane. We discard any 2D slice where at least one of the ROIs contains less than 250 pixels. This is because:

(a) When the tumor is small, the main challenge becomes object detection, not segmentation.
(b) Slices with extremely small tumors are more sensitive to incorrect labeling from an annotator. The Dice score punishes a single-pixel mistake more heavily for small tumors than for larger tumors.

### 3.8. Preprocessing:

The Decathlon data represents each brain as a 4D volume (155 x 240 x 240 brain and 4 modalities). We sliced along the axial plane to generate 155 3D images (240 x 240 x 4) per brain. Then we normalized the intensities by clipping the 0.5 and 99.5 percentiles of intensity and then linearly mapping the intensities to the range [0, 255].

SAM requires three channels (RGB). We used three methods for selecting the three channels.

**Method 1:** We chose three distinct modalities and put one modality in each channel. We decided to use FLAIR, T1-Gd, and T2w, in that order. Out of the six combinations we tried, using FLAIR, T1-Gd, and T2w images together gave slightly better results than the others.

**Method 2:** We chose a single modality and repeated it for all three channels. (T1-Gd gave slightly better results overall)

**Method 3:** We used 2D-PCA to reduce the four channels to three channels.

Finally, all the 3D images were resized to 1024 x 1024 x 3 high resolution as the model requirement.

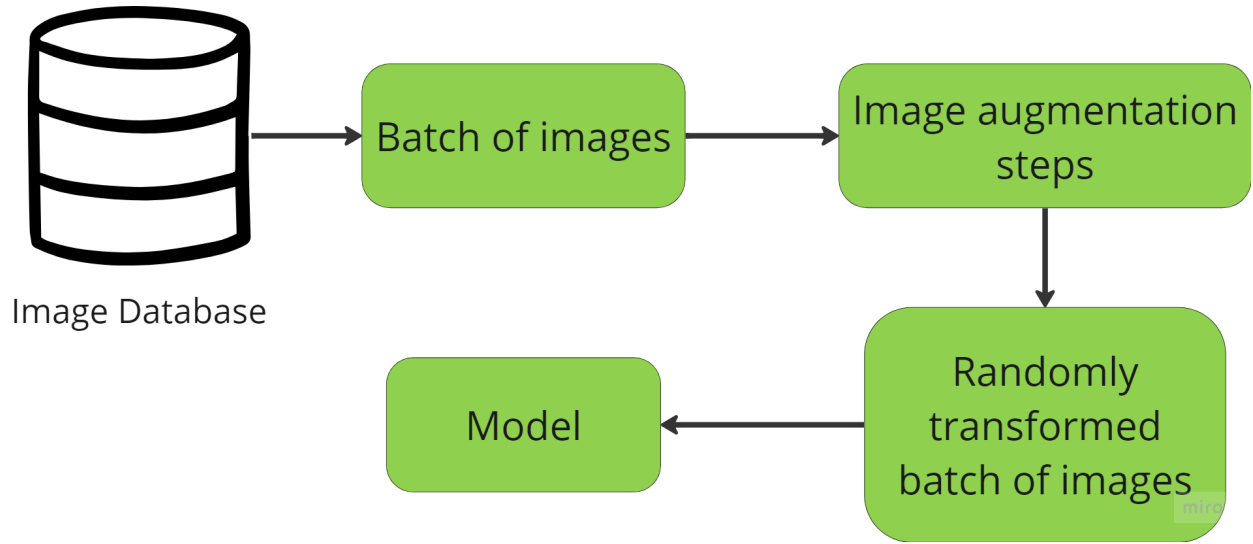

Fig 1: A diagram illustrating the image augmentation pipeline. A batch of images is taken from the image database. Then image augmentation is randomly applied "on-the-fly". Finally, the transformed batch of images is fed to the model for training.

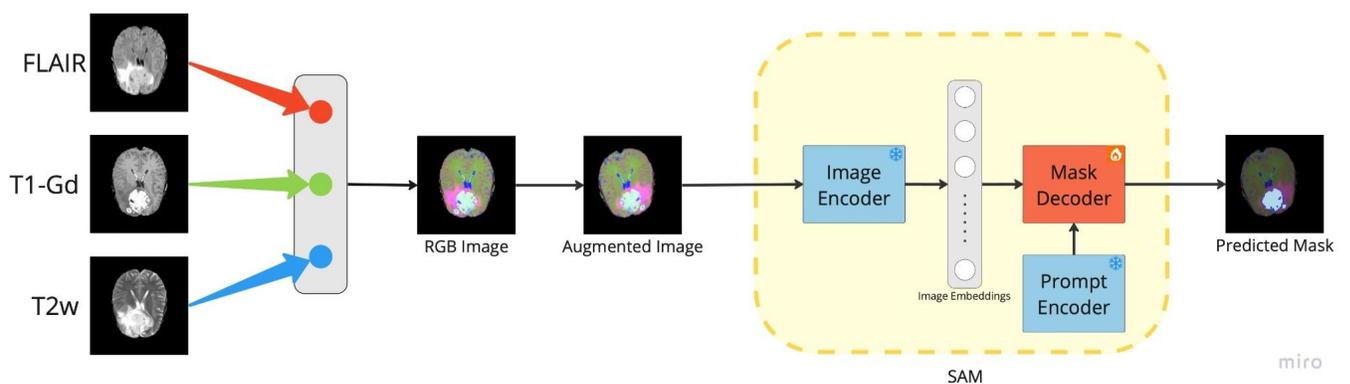

Fig 2: A diagram showing the training process of the proposed model. Here, three separate modalities (FLAIR, T1-Gd, T2w) are fed into the red, green, and blue channels, respectively, of SAM's image encoder. Both the image encoder and prompt encoder parameters are frozen, so that only the mask decoder is trained. The output is the predicted mask.

## 4. Results:

In this paper, we investigated the potential of the fine-tuned version of SAM on brain tumors. In doing so, we used three methods of channeling MRI modalities into the RGB requirements of the pretrained SAM. This involved experimenting with six different combinations of MRI modalities(Table 2), employing a PCA-driven representation where the dimensionality reduction technique of early fusion 2D-PCA was used to transform the four MRI modalities into three RGB channels, and finally by using the repetitive modality method which adopted the T1-Gd modality across all three channels (Table 1). 2D-PCA method showed a lower Dice score in each ROI even though using four modalities (Table 3).

Table 1. Dice Score of T1-GD Repeated method

| T1-Gd Repeated | Pretrained SAM | Improved Model | nnUNetv2 |
|---|---|---|---|
| Enhancing | 0.60 | 0.80 | 0.83 |
| Non-enhancing | 0.37 | 0.56 | 0.61 |
| Edema | 0.32 | 0.55 | 0.65 |
| Whole Tumor | 0.68 | 0.80 | 0.85 |

Table 2. Dice Score of the combined method*.

* Combined refers to loading the modalities FLAIR, T1-Gd, and T2w into the RGB channel input, in that order.

| Combined* | Pretrained SAM | Improved Model | nnUNetv2 |
|---|---|---|---|
| Enhancing | 0.53 | 0.75 | 0.84 |
| Non-enhancing | 0.38 | 0.54 | 0.64 |
| Edema | 0.41 | 0.67 | 0.78 |
| Whole Tumor | 0.75 | 0.88 | 0.93 |

Table 3. 2D-PCA Dice Score

| Method | Modalities | Enhancing | Non-enhancing | Edema | Whole Tumor |
|---|---|---|---|---|---|
| 2D-PCA | Flair, T1, T1gd, T2 | 0.71 | 0.53 | 0.55 | 0.78 |

## Enhancing

** ***

Pretrained SAM | Improved Model | nnUNetv2

## Non-enhancing

** ***

Pretrained SAM | Improved Model | nnUNetv2

## Edema

** ***

Pretrained SAM | Improved Model | nnUNetv2

## Whole Tumor

** ***

Pretrained SAM | Improved Model | nnUNetv2

Fig 3: Box plots of Dice score on our test dataset.

\* Combined refers to loading the modalities FLAIR, T1-Gd, and T2w into the RGB channel input, in that order.

\*\* The Wilcoxon signed-rank test shows that the fine-tuned version of SAM (Improved Model) is significantly better than the pretrained SAM ($p<0.001$)

\*\*\* The Wilcoxon signed-rank test shows that nnUNetv2 is better than the improved model ($p<0.001$)

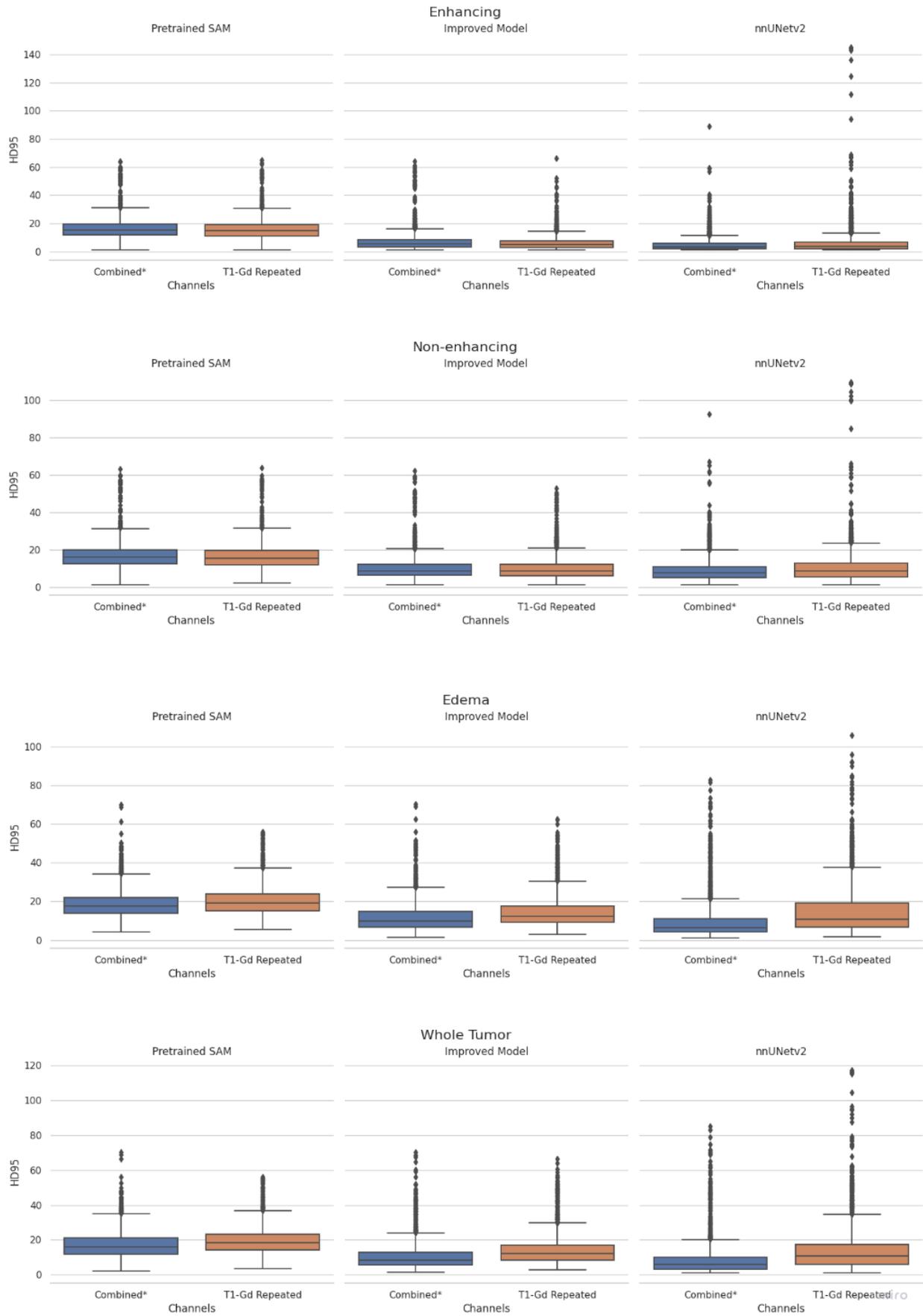

Fig 4: Box plots of HD95 on our test dataset.

* Combined refers to loading the modalities FLAIR, T1-Gd, and T2w into the RGB channel input, in that order.

## 5. Discussion:

Even though HD95 uses the 95th percentile, there are still many bad segmented images in both the nnUNetv2 and the improved model. In Figure 5 we see these examples that show a slightly lower consistency in nnUNetv2. In Figure 6 we see an uncommon case where the improved model does better.

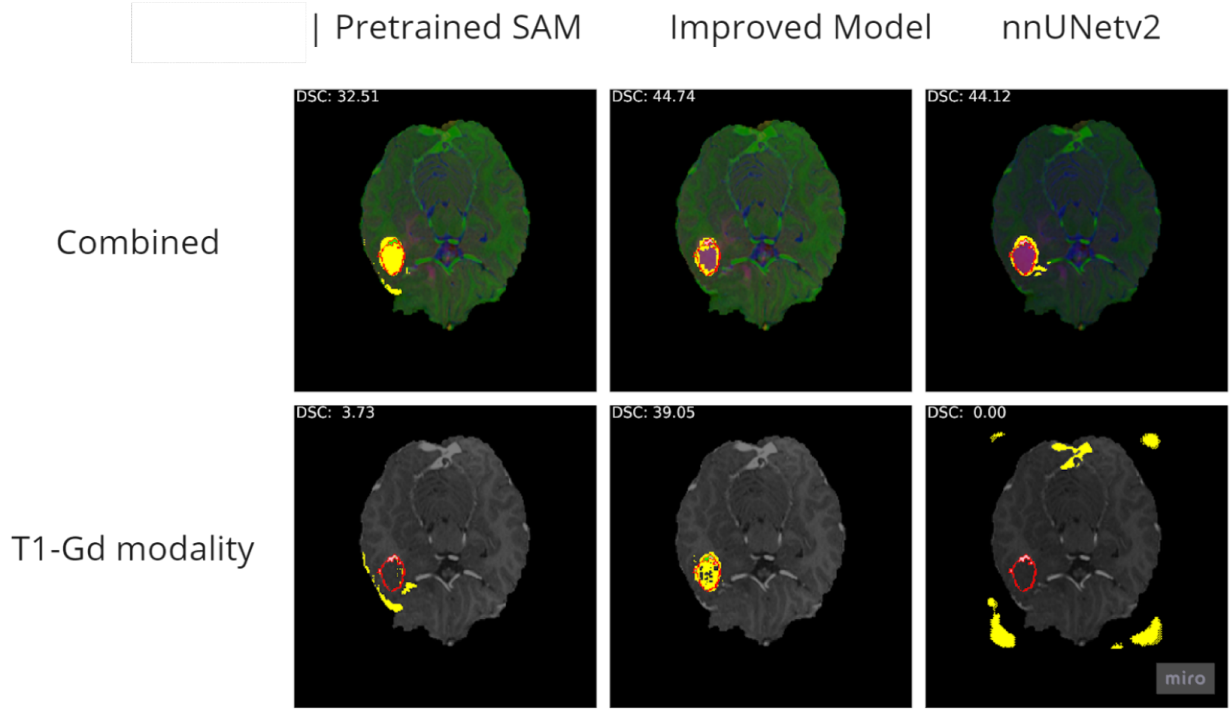

Fig 5: Very poor performance of some nnUNetv2 segmentation compared to the Improved model. The green region represents pixels that the model correctly predicted as true. The yellow region represents pixels that the model incorrectly predicted as true. The pink region represents ground truth pixels which the model incorrectly predicted as false. The red boundary represents the border of the ground truth and does not correspond to the model prediction.

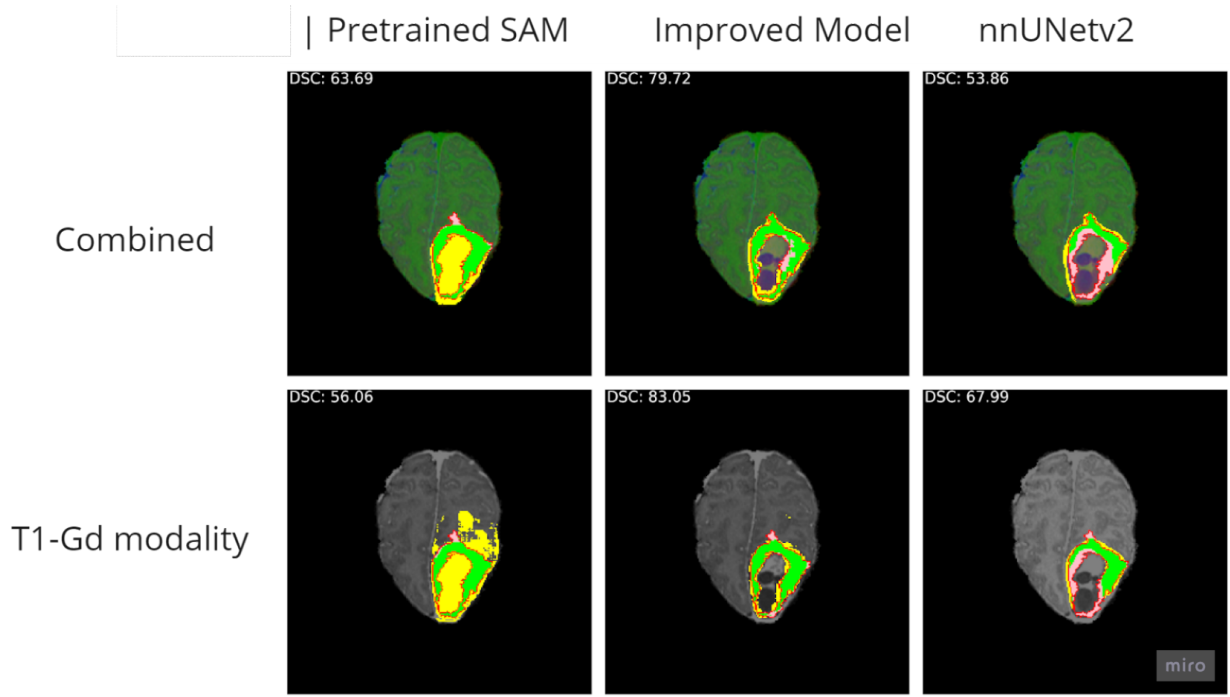

Fig 6: Examples of segmented tumors. The background image is darkened to highlight the ground truth and model prediction. The green region represents pixels that the model correctly predicted as true. The yellow region represents pixels that the model incorrectly predicted as true. The pink region represents ground truth pixels which the model incorrectly predicted as false. The red boundary represents the border of the ground truth and does not correspond to the model prediction.

There are many outliers in the Dice scores due to slices such as the figures above or Figure 8. Since we use a bounding box that covers the whole ground truth as the prompt for the improved model, the improved model may not recognize two or more completely separate regions as unconnected. Thus, while the output may include the ground truth, it will try to predict part of the brain between the separated ground truth regions in its segmentation as well, or just ignore the separated ground truths entirely, resulting in a low Dice score. Figure 7 shows a frequent case in which the improved model and nnUNetv2 beat the pretrained SAM model.

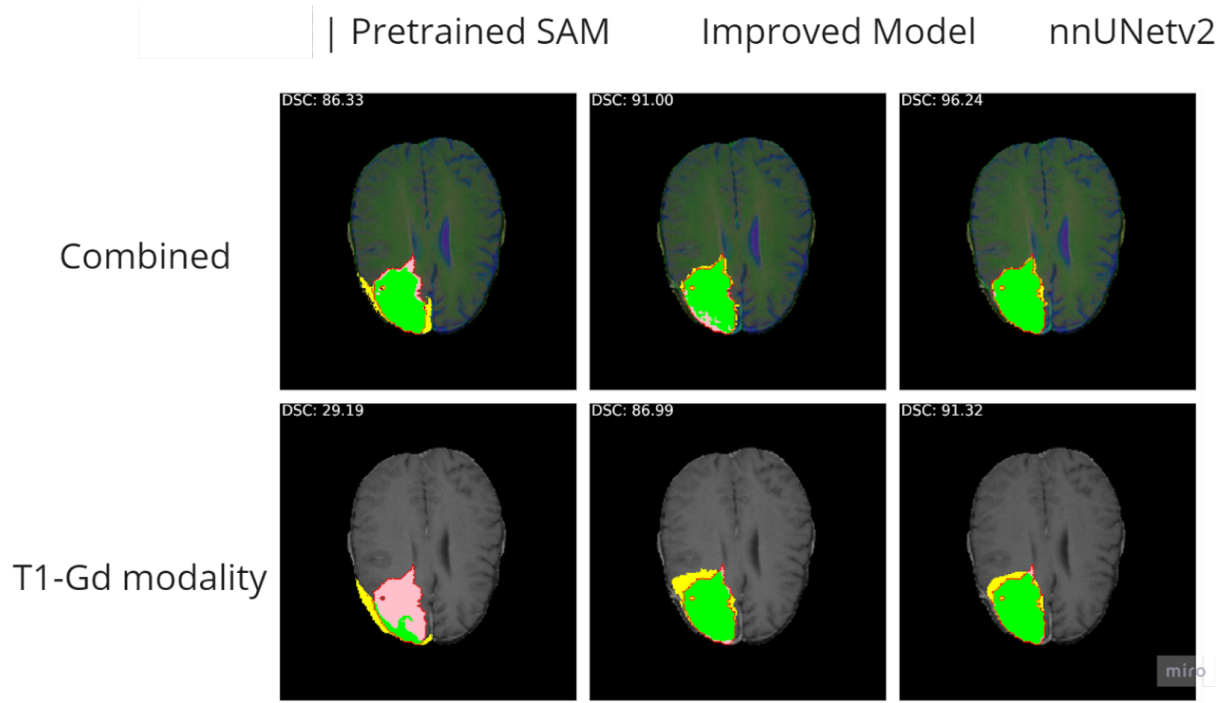

Fig 7: Examples of segmented tumors that show the poor performance of pretrained SAM. Ground truths such as the one above result in very low dice scores due to the abnormal shape of the tumor. The pretrained model tries to output a whole region as the mask while our proposed model outputs a few disconnected regions. These tumors are extremely difficult to segment properly even for trained experts, so the lower scores are not surprising.

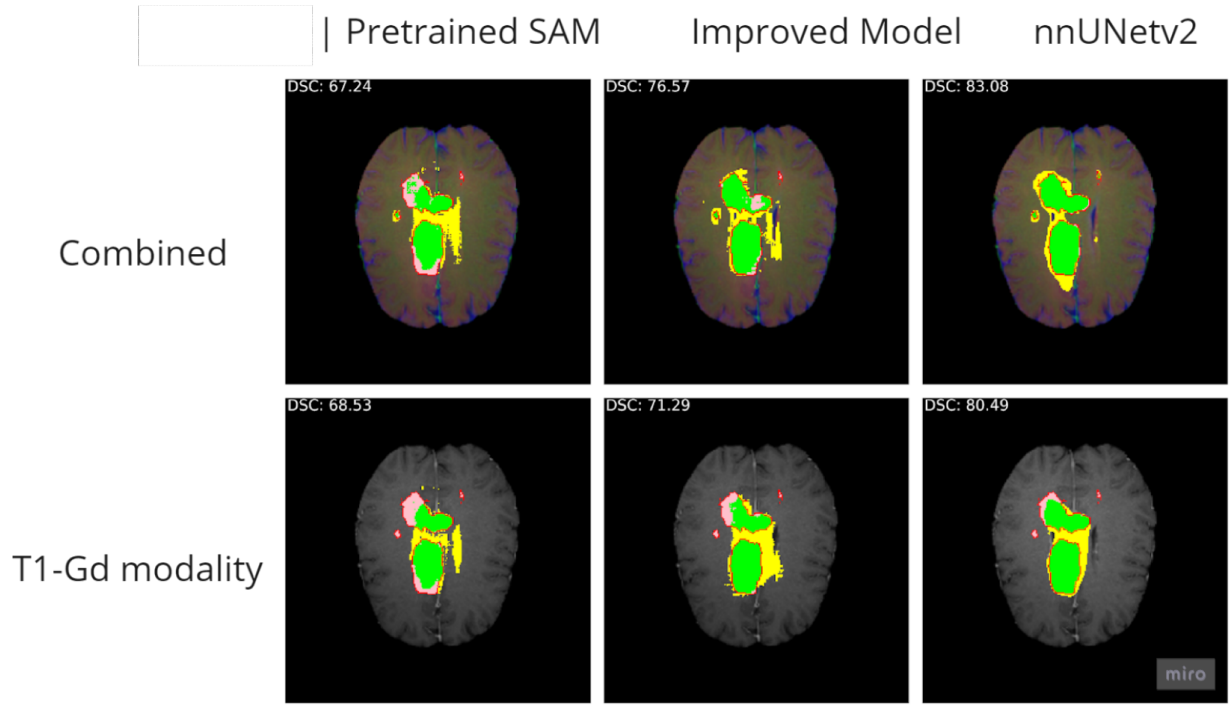

Fig 8: Lower segmentation results for separated ground truth regions. The green region represents pixels that the model correctly predicted as true. The yellow region represents pixels that the model incorrectly predicted as true. The pink region represents ground truth pixels which the model incorrectly predicted as false. The red boundary represents the border of the ground truth and does not correspond to the model prediction.

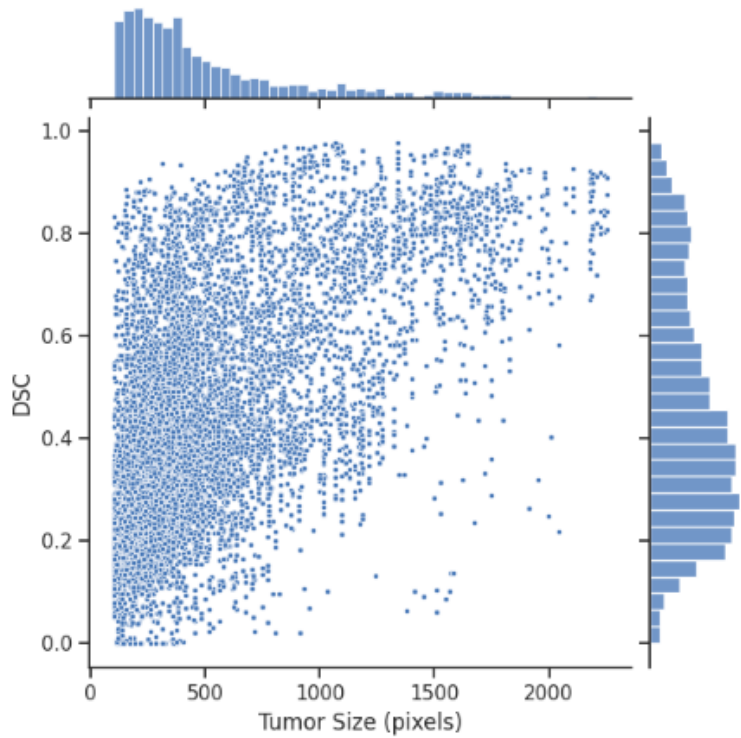

Fig 9: A scatterplot showing the relation between the Tumor size and Dice score. Larger tumor sizes resulted in higher minimum DSC than smaller tumor sizes.

Data augmentation techniques helped slightly with the overfitting happening in the training. The SAM has a huge number of parameters (91 Million parameters) relative to the number of training data samples, which makes data augmentation challenging.

Fine-tuning the image encoder requires a very high computational power. SAM has a strong image encoder and can extract the features even from medical images very well. So freezing the image encoder is a good approach to save time and computational power.

Using the base model of SAM instead of huge or large was due to the fact that they do not increase the performance as much while they are a much larger model. [14]

RGB channels permutations were not very much different; the model cannot make full use of different modalities, meaning it cannot distinguish between different modalities.

The SAM model, pretrained on natural images, has biases towards RGB visual patterns found in everyday objects and scenes. When using three modalities, the transformed representation might not align well with the model's pretrained features, leading to lower performance on some subregions.

## 6. Conclusion:

In this study, we fine-tuned the SAM mask decoder on the Decathlon brain tumor dataset. The Decathlon brain tumor dataset comprises multimodal 3D volumes with necrotic/active gliomas and edema. We compared the performances of the improved model with that of two other models: the widely usedt model, nnUNetv2, and pretrained SAM.

The improved SAM was better than pretrained SAM, but was still lacking compared to nnUNetv2. The pretrained SAM was unable to segment complex boundaries, especially for more challenging ROIs such as non-enhancing lesions and edema. In contrast, the improved model showed improvement over pretrained SAM, but still missed on complex or discrete regions which nnUNetv2 was slightly better at.

Even though nnUNetv2 mainly performs better than the improved model, from Figure 4, we see the box plot of HD95 values shows a higher spread and presence of outliers. The nnUNetv2 model might occasionally produce segmentation results with significant boundary discrepancies. This shows the importance of multiple evaluation metrics. While nnUnetv2 is better in general segmentation accuracy, the improved SAM model offers slightly more consistent results, especially in challenging cases that can lead to larger Hausdorff distances.

The multimodal method (3 modalities) showed comparable performance to the repeating channel method, which indicates the necessity of more advanced methods, such as advanced fusion techniques, to fully utilize all four modalities.

Furthermore, our results showcase the difference in segmentation performance between ROIs. The non-enhancing ROI was the most challenging for both our model and nnUNetv2. In brain tumor segmentation, the non-enhancing tumor region can exhibit highly heterogeneous and irregular shapes, making it complex to identify. After the administration of a contrast agent, while the enhancing tumor segments become more apparent, the non-enhancing tumor does not show this change. This lack of differentiation post-contrast can lead to confusion between the non-enhancing tumor and other regions, especially edema.

The segmentation of edema and enhancing ROI demonstrated better results due to their well-defined borders, more regular shapes, pronounced contrast dynamics after the administration of a contrast agent, and the distinct image properties associated with edema such as swelling and fluid accumulation.

In conclusion, the improved model showed promising performance on brain tumor segmentation, but other adjustments are needed in order to be made comparable to widely used methods. Our findings can help future research on SAM's capability to segment tumors.